\documentclass[a4paper]{jpconf}

\usepackage{graphicx}
\usepackage{amsmath}
\usepackage{wrapfig}
\usepackage{url}

\begin{document}
\title{A development of an accelerator board dedicated for 
       multi-precision arithmetic operations and its application to Feynman loop integrals}

\author{
S~Motoki$^1$,
H~Daisaka$^2$,
N~Nakasato$^3$,
T~Ishikawa$^1$,
F~Yuasa$^1$,
T~Fukushige$^4$,
A~Kawai$^4$,
J~Makino$^5$
}

\address{$^1$ High Energy Accelerator Research Organization (KEK), 1--1, Oho, Tsukuba, Ibaraki, 305--0801, Japan}
\address{$^2$ Hitotsubashi University, 2--1, Naka, Kunitachi, Tokyo, 186--0801, Japan}
\address{$^3$ University of Aizu, Aizu-wakamatsu, Fukushima,  965--8580, Japan}
\address{$^4$ K\&F Computing Research Co., 1--21--6--407, Kojimacho, Chofu, Tokyo, 182--0026, Japan}
\address{$^5$ RIKEN Advanced Institute for Computational Science, 7--1--26, Minatojima-minami-machi, Chuo-ku, Kobe, Hyogo, 650--0047, Japan}

\ead{smotoki@post.kek.jp}

%%%%%%%%%%%%%%%%%%%%%%%%%%%%%%%%%%%%%%%%%%%%%%%%%%%%%%%%%%%%%%%%%%%%%%%%%%%%%%%%%%%%%%
\begin{abstract}
Higher order corrections in perturbative quantum field theory are required 
for precise theoretical analysis to investigate new physics beyond 
the Standard Model. 
This indicates that we need to evaluate Feynman loop diagrams
with multi-loop integrals which may require multi-precision calculation.
We developed a dedicated accelerator system
for multi-precision calculations (GRAPE9-MPX).
We present performance results of our system for the case of Feynman two-loop box 
and three-loop selfenergy diagrams with multi-precision. 
\end{abstract}
%
%
%
%%%%%%%%%%%%%%%%%%%%%%%%%%%%%%%%%%%%%%%%%%%%%%%%%%%%%%%%%%%%%%%%%%%%%%%%%%%%%%%%%%%%%%
\section{Introduction}
With the discovery of Higgs particle at the CERN Large Hadron Collider, 
precision measurements are expected at a future International Linear Collider 
to explore new physics beyond the Standard Model. 
In tandem with experiments, an accurate theoretical prediction is required.
To meet such a demand, higher order correction in perturbative quantum field theory 
becomes more and more important, and methods to evaluate 
multi-loop integrals precisely should be provided. 

We have been developing DCM (Direct Computation Method) 
for loop integrals \cite{Yuasa_2004}. 
This is a fully numerical method of the combination of multi-dimensional integration 
and the extrapolation technique. 
It is known that the accurate numerical evaluation of the integral is 
a hard problem due to its divergent nature.
Yuasa et al. \cite{Yuasa_2007} reported that the numerical evaluation is 
numerically unstable with double-precision operations. 
A solution to this difficulty is to compute the integral in multi-precision arithmetic,
that is, it needs more bits for mantissa than double-precision.
In addition, we have a case that needs more bits for exponent.
With Double Exponential Formulas for numerical integration (DE) \cite{Mori_2005},
we can estimate that 15-bit wise exponent is required to reduce a relative 
error smaller than a reference value,  
although the exponent of double-precision is only 11-bit wise. 

%%%%%%%%%%%%%%%%%%%%%%%%%%%%%%%%%%%%%
There are several ways to accomplish multi-precision arithmetic. 
One way is to use the specific softwares, for examples, 
{\tt GMP} \cite{GMP}, {\tt MPFR} \cite{MPFR}, 
{\tt ARPREC} \cite{QDLIB}, {\tt MPFUN90} \cite{QDLIB}, {\tt DD/QD} \cite{QDLIB}, 
and {\tt quadmath} 
on a new version of {\tt GNU} compiler.
These softwares are easy to use.
%However, 
%high performance can not be expected.
%Another way is to design a new hardware 
%which has a dedicated logic for calculation with precision higher 
%than double-precision.
However, 
high performance can not be expected, 
since one multi-precision arithmetic operation is realized with,  
at least, more than 10 double-precision arithmetic operations in these softwares.
Another way is to design a new, dedicated hardware 
which has circuits for calculation with precision higher 
than double-precision.
We have been developing such a hardware 
based on the fundamental idea of GRAPE  
originally developed for gravitational $N$-body simulations \cite{Koike_2009}. 
We designed GRAPE-Multi-Precision (MP) processor which consisted of a number of 
%processing elements (PE) and memory components with a dedicated logic 
processing elements (PE) and memory components with dedicated circuits
for quadruple, hexuple, octuple-precision arithmetic.
We implemented this processor on a structured ASIC (Application Specific Integrated Circuit)
which was called GRAPE-MP \cite{Daisaka_2011} , 
and on an FPGA (Field Programmable Gate Array) board with a control processor, 
which was called GRAPE-MPX \cite{Nakasato_2012}.
On these systems, 
we measured the performance of quadruple, hexuple, octuple-precision calculation 
and obtained the higher performance compared to a software implementation.

In this work, we developed a new system called GRAPE9-MPX. 
It is similar to GRAPE-MPX but is a rather larger system.
This means that larger number of PEs can be implemented in the new system,
which will be suited for large scale simulations.
In addition to designing a hardware, 
we have been developing a programming interface named Goose and LSUMP 
which enable us easily to use GRAPE9-MPX.

This paper is organized as follows. 
Next section covers a brief explanation of the design and the implementation
of the new system from hardware and software viewpoints.
In section 3, we briefly explain Feynman loop integrals used for the performance 
measurement. The results are shown in section 4.
The last section is devoted to the summary and our future prospects.
%
%
%
%%%%%%%%%%%%%%%%%%%%%%%%%%%%%%%%%%%%%%%%%%%%%%%%%%%%%%%%%%%%%%%%%%%%%%%%%%%%%%%%%%%%%%
\section{GRAPE9-MPX system}
\label{sec:g9mpx-sys}

\subsection{Hardware}

%%%%%%%%%%%%%%%%%%%%%%%%%%%%%%%%%%%%%%%%%
\begin{figure}[htbp]
 \begin{minipage}{0.5\hsize}
  \begin{center}
   \includegraphics[width=80mm]{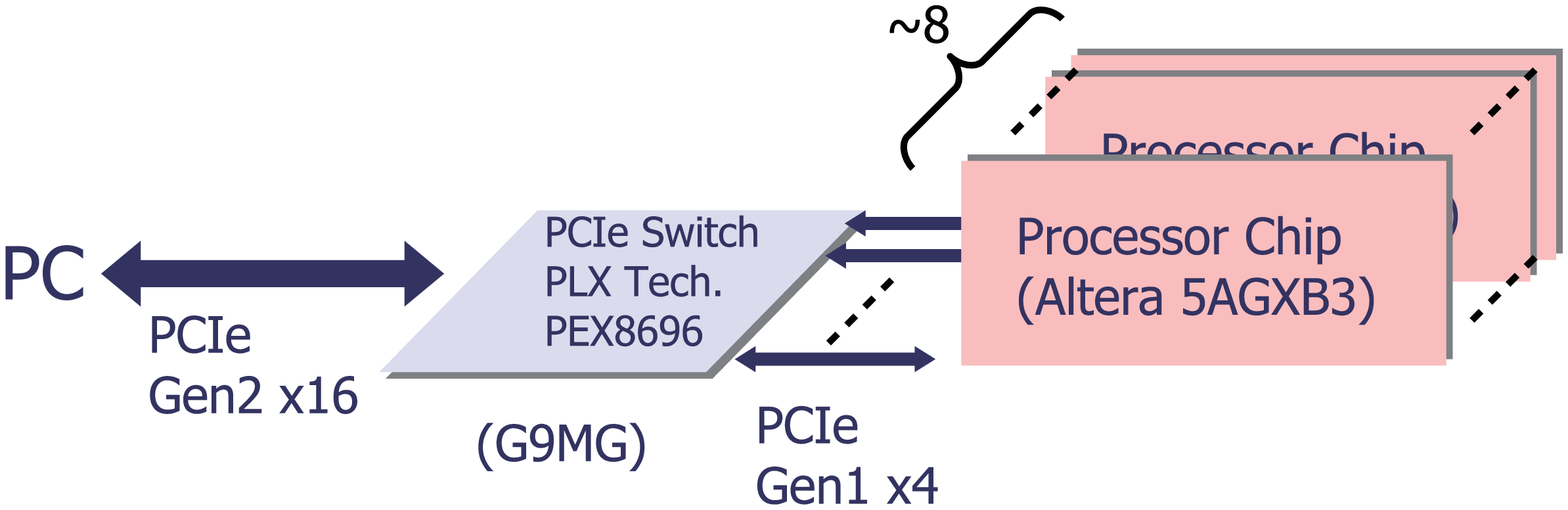}
  \end{center}
  %\caption{}
  %\label{fig:imag-g9mpx}
 \end{minipage}
 \begin{minipage}{0.5\hsize}
  \begin{center}
   \includegraphics[width=50mm]{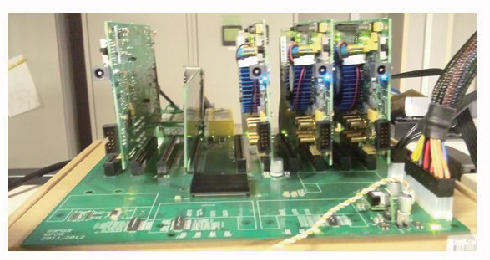}
  \end{center}
  %\caption{}
  %\label{fig:pict-g9mpx}
 \end{minipage}
  \caption{Schematic diagram (left) and picture (right) of GRAPE9-MPX.
  Note that 4 FPGA boards are included in the system.}
  \label{fig:imag-g9mpx}
\end{figure}
%%%%%%%%%%%%%%%%%%%%%%%%%%%%%%%%%%%%%%%%%

Figure \ref{fig:imag-g9mpx} shows a schematic diagram and a picture of GRAPE9-MPX system.
It is an accelerator which is connected to a host computer.
It consists of FPGA boards (up to 8 boards available) which are housed
on a PCIe extender board (G9MG).
For the FPGA board, we used Altera Arria V board from Altera Co. \cite{AlteraArriaVboard}.
On this FPGA, we implemented the MP processor and its Control Processors (CP). 
The MP processor consists of PEs and memory components 
for quadruple/hexuple/octuple-precision arithmetic (named as MP4/MP6/MP8),  
which forms a Single-Instruction-Multiple-Data (SIMD) processor.
Each PE has a floating-point multiply unit and an add unit 
which perform in every clock cycle.
Note that for MP4, 
we used the numerical representation compatible with IEEE-754-2008:binary128 format. 
For details of the PE, CP, and the numerical representations, see
Daisaka et al. \cite{Daisaka_2011} and Nakasato et al. \cite{Nakasato_2012}.

One of applications suited for our system is an interaction type calculation as
$f_{i} = \sum_{j=1}^{n_{j}} f(X_i,  Y_j), $
where $X_{i}$ is $i$-th element of $X$, 
$Y_{j}$ is $j$-th element of $Y$, 
and $n_{j}$ is the number of elements of $Y$.
The function $f(X_{i}, Y_{j})$ describes an interaction form of $X_{i}$ and $Y_{j}$.
Note that data $X_{i}$ is set on $i$-th PE,
whereas data $Y_{j}$ is set on all PE,
then each PE calculates $f(X_{i}, Y_{j})$ and sums up from $j=1$ to $j=n_j$ 
in accordance with instructions. 
An example of this type calculation is gravitational interactions among particles.
We should note that a multi-dimensional integration can be expressed in 
a similar form by {\it loop fusion} technique that merges multiple loops 
into a single loop.
This indicates that we can accelerate the calculation of 
Feynman loop integrals effectively by using GRAPE9-MPX.

Table \ref{logic_utilization} summarizes our current implementation of 
MP4/MP6/MP8 processors on the FPGA board. 
The peak performance of MP4, for example, can be estimated as
2 (operations) $\times$ 36 (PEs) $\times$ 92 (MHz) = 6.6 Gflops per board. 
It adds up to  26.4 Gflops for the system with 4 boards.

%%%%%%%%%%%%%%%%%%%%%%%%%%%%%%%%%%%%%%%%%
\begin{table}[htb]
\begin{center}
\begin{tabular}{llll} \hline
{}                       &  MP4   &  MP6   & MP8 \\ \hline
Number of PEs            &  36    &  20    & 12 \\
Clock Speed(MHz)         &  92    &  81    & 70 \\
Theoretical Peak(Gflops) &  6.6   &  3.2   & 1.6 \\ \hline
\end{tabular}
\end{center}
\caption{
Number of PEs, clock speed, and theoretical peak performance for MP4/MP6/MP8 
implemented on the FPGA.}
\label{logic_utilization}
\end{table}

\subsection{Programming interface}

\begin{figure}[htbp]
 \begin{minipage}{0.5\hsize}
  \begin{center}
   \includegraphics[width=4.5cm]{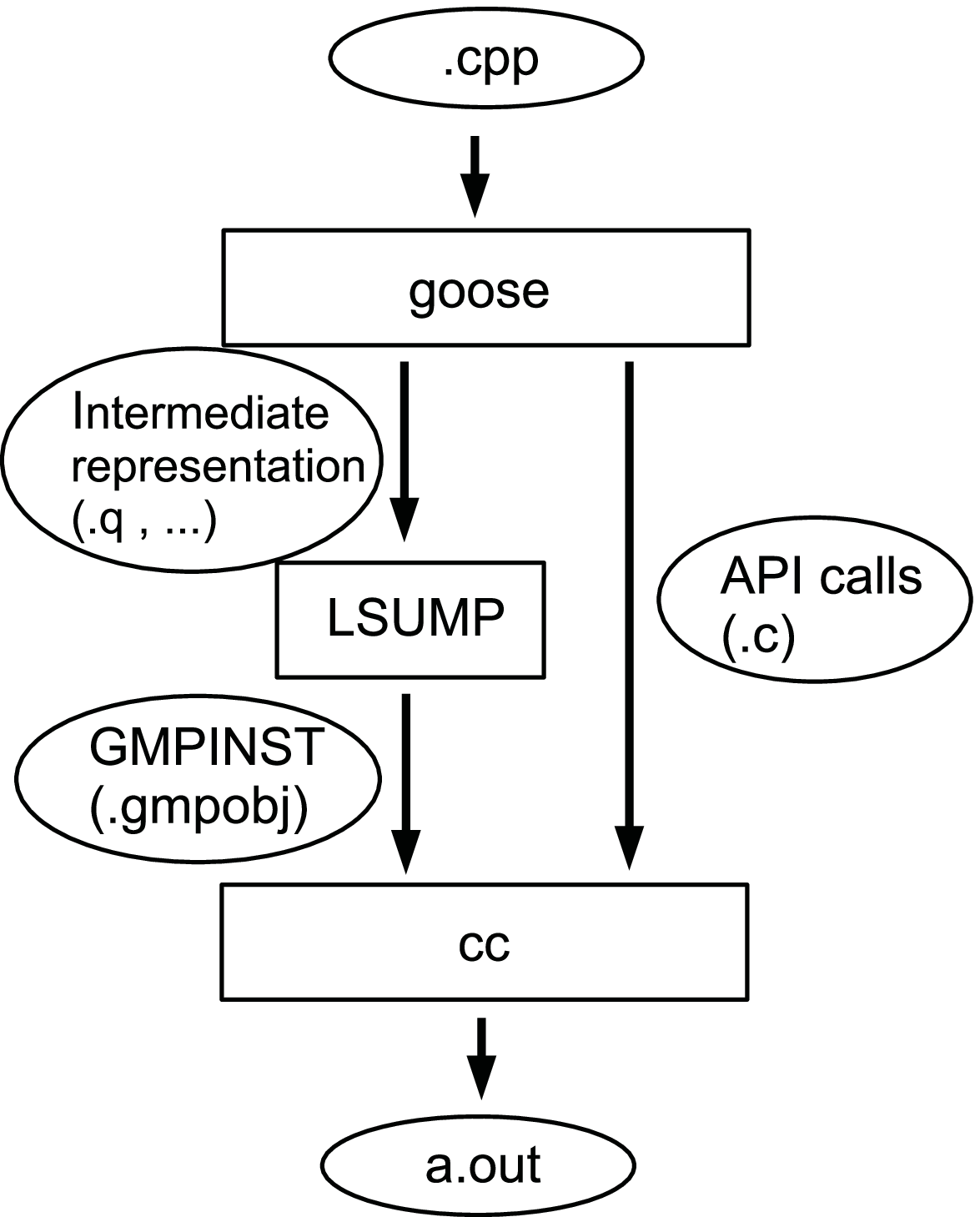}
%%\caption{Loop operation off-road programming by Goose }
  \end{center}
  %\caption{}
  %\label{fig:imag-g9mpx}
 \end{minipage}
 \begin{minipage}{0.5\hsize}
  \begin{center}
   \includegraphics[width=8.0cm]{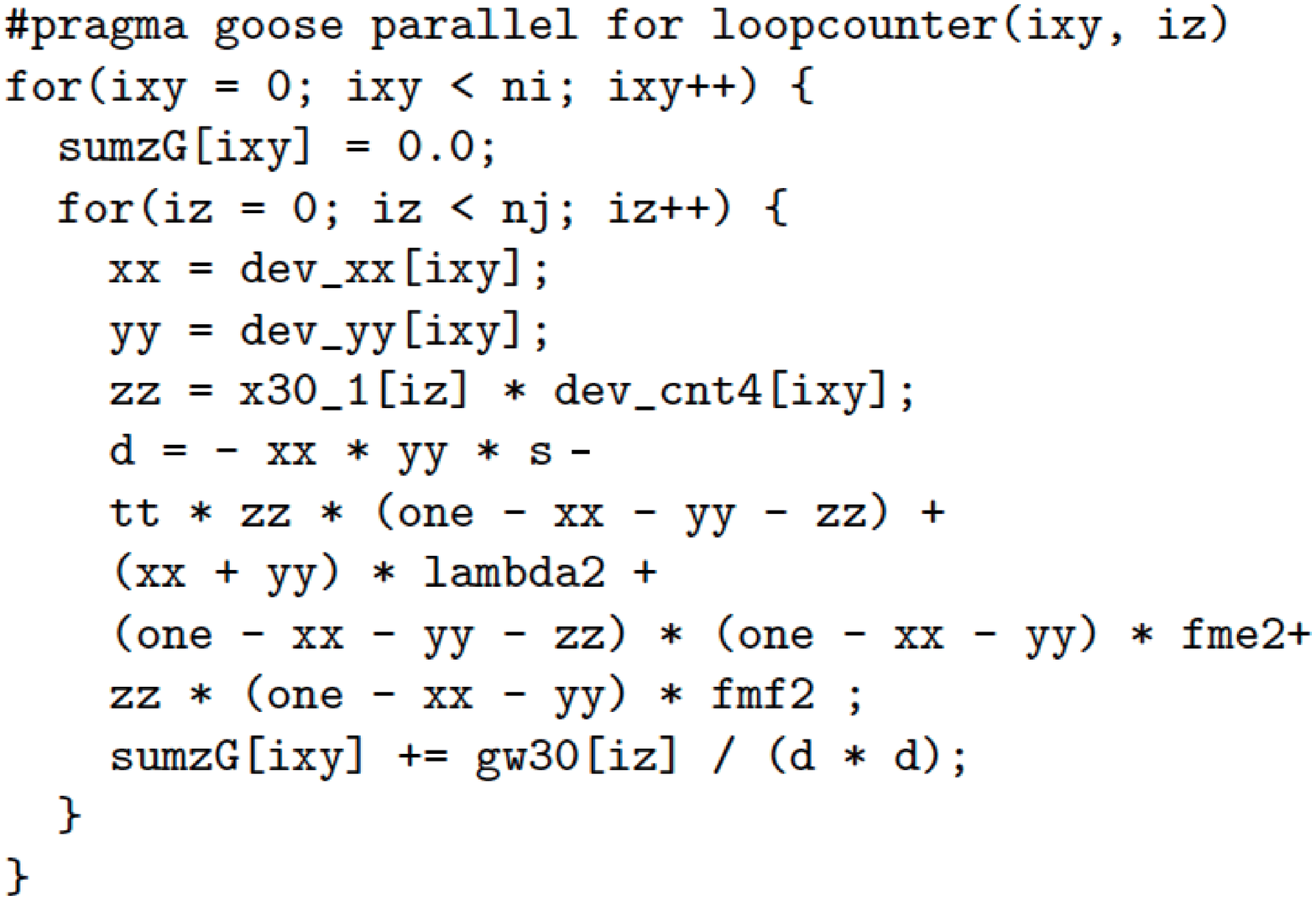}
  \end{center}
  %\caption{}
  %\label{fig:pict-g9mpx}
 \end{minipage}
%%  \caption{schematic image(left) and picture(right) of GRAPE9-MPX}
\caption{Flow of the compile process (left) and 
 a sample code with the directive (right) for our system.}
  \label{fig:gooseflow}
\end{figure}

We have also developed Goose compiler which provides 
a programming interface for GRAPE9-MPX system.
Figure \ref{fig:gooseflow} shows the flow of the compile process for our system, 
and the part of a sample program for one-loop box integral.
Goose is a directive base compiler like OpenMP,
in which a directive is inserted in an original code. 
By the directive (\textsc{\#pragma goose parallel} seen in the sample code), 
Goose extracts the loop next to the directive 
and generates an intermediate representation which is like an assembler code.
Goose also generates API calls in which functions of I/O interfaces 
for GRAPE9-MPX are embedded.
In order to convert the intermediate representation to machine instructions (kernel code) 
for the MP processor, 
Goose uses LSUMP backend \cite{Nakasato_2009} which is a domain specific language (DSL) compiler
specially developed for the MP processor. 
The kernel code is read by an executable file which is generated from the API calls 
by C/C++ compiler.
For more detail, see Nakasato \cite{Nakasato_2009}.

%%%%%%%%%%%%%%%%%%%%%%%%%%%%%%%%%%%%%%%%%%%%%%%%%%%%%%%%
%\begin{table}[htp]
%\begin{center}
%  \begin{tabular}{rrrrrr} \hline
%          & MP4 & MP6 & MP8 & DD  & QD \\ \hline
%total bit & 128 & 192 & 256 & 128 & 256 \\
%sign      &   1 &   1 &   1 &   1 &   1 \\
%exponents &  15 &  15 &  15 &  11 &  11 \\
%mantissa  & 112 & 176 & 240 & 106 & 212 \\\hline
%  \end{tabular}
%\caption{Numeric representation which we deal with by GRAPE9-MPX }
%\label{tab:numerical_representation}
%\end{center}
%\end{table}
%%%%%%%%%%%%%%%%%%%%%%%%%%%%%%%%%%%%%%%%%%%%%%%%%%%%%%%%%%
%
%%%%%%%%%%%%%%%%%%%%%%%%%%%%%%%%%%%%%%%%%
%%%%%%%%%%%%%%%%%%%%%%%%%%%%%%%%%%%%%%%%%
%
%
%
%%%%%%%%%%%%%%%%%%%%%%%%%%%%%%%%%%%%%%%%%%%%%%%%%%%%%%%%%%%%%%%%%%%%%%%%%%%%%%%%%%%%%%
\section{Feynman loop integrals}
\begin{figure}[htbp]
 \begin{minipage}{0.5\hsize}
  \begin{center}
   \includegraphics[width=40mm]{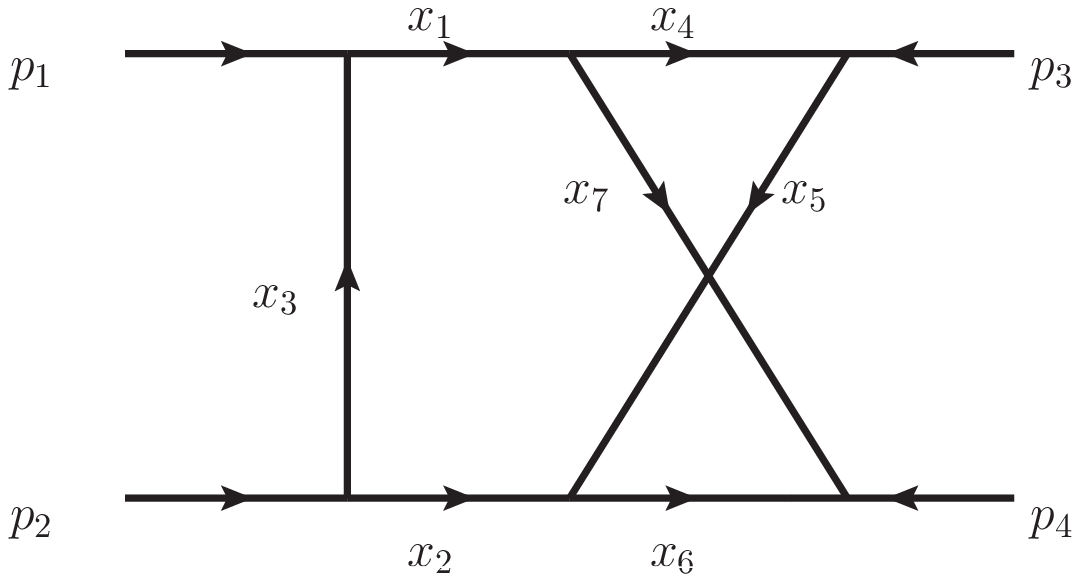}
  \end{center}
%  \caption{Two-loop crossed box diagram}
%  \label{fig:2lbc}
 \end{minipage}
\hfill
 \begin{minipage}{0.5\hsize}
  \begin{center}
   \includegraphics[width=35mm]{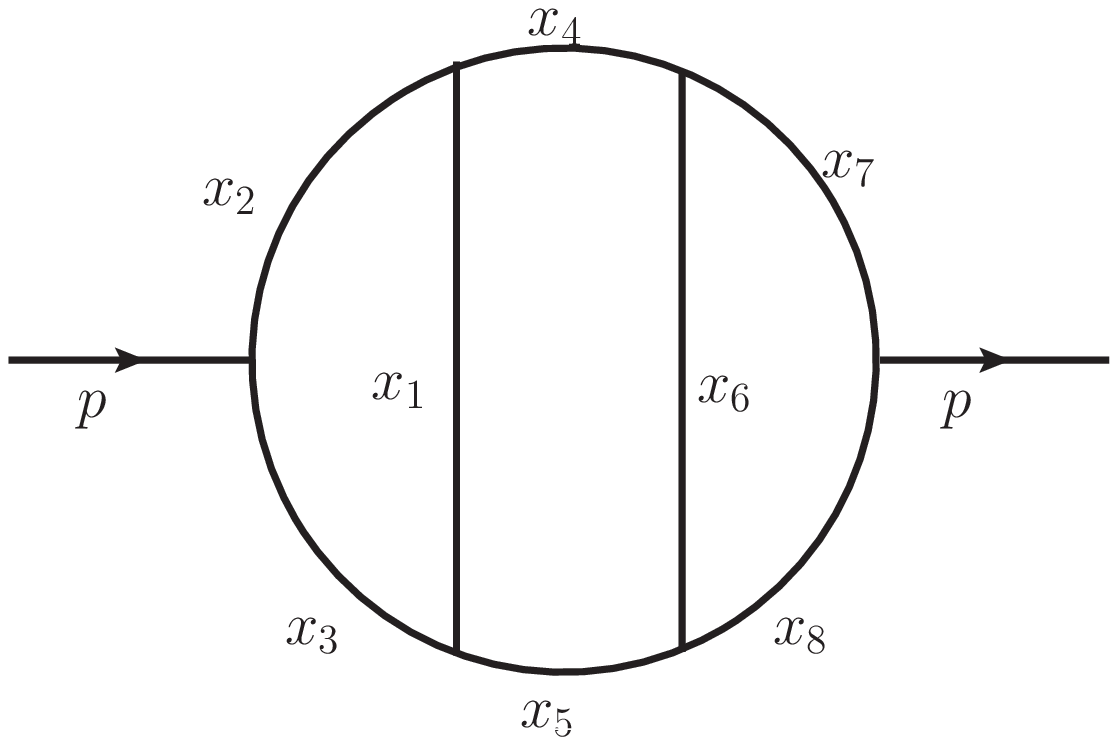}
  \end{center}
%  \caption{ Three-loop selfenergy diagram}
%  \label{fig:3ls}
 \end{minipage}
  \caption{ Two-loop crossed box diagram and three-loop selfenergy diagram used for the
  performance measurement. In these diagrams, $p_i$ and $x_i$ denote 
a momentum of an external line and a Feynman parameter, respectively. }
 \label{fig:2lbc-3ls}
\end{figure}

Figure \ref{fig:2lbc-3ls} shows the two-loop crossed box diagram and the three-loop 
selfenergy diagram used for the performance measurement.
%%%%%%%%%%
%%%%%%%%%%
The number of dimensions of the loop integral is 
six for the two-loop crossed box, 
and seven for the three-loop selfenergy, respectively.
We consider the case where all masses of the internal lines are 1 and set the Mandelstam variables $s=t=1$ and the external momenta $p_i^2=1 (1\le i \le 4)$ for the two-loop crossed box diagram and $s=1$ and $p^2=1$ for the three-loop selfenergy diagram, respectively. We choose these values to compare numerical results to ones in the previous studies \cite{Yuasa_2012,Laporta_2000}.

%We consider a simple case where all masses of the internal lines and 
%%the Mandelstam variables are set to 1, so that 
%the kinematical variables are set to 1, so that 
%there is no threshold 
%divergence inside the integration domain except at the endpoint.

With these parameters, 
the integrals of the two-loop crossed box with variable transformations from $\{x_i\}$ to $\{\xi_i\}$
to change the integration domain to the unit cube from the unit simplex is expressed as 
\begin{eqnarray}\label{eq:I2lbc}
I_{2loop} &=& 2 \int_{0}^{1} d\xi_{1} ~d\xi_{2} ~d\xi_{3} ~d\xi_{4} ~d\xi_{5} ~d\xi_{6} 
 ~(\xi_1^2 {\xi'_1}^3 \xi'_2 \xi_4 \xi'_4 ) \frac{C}{D^{3}}, \\ \nonumber
\end{eqnarray}
where 
\begin{eqnarray}\label{eq:I2lbc2}
C &=& \xi'_1 (\xi_1+ \xi'_1 \xi_4 \xi'_4), \\ \nonumber
D &=& \xi'_1 ((\xi_1+ \xi'_1 \xi_4 \xi'_4) \\ \nonumber
&-&(
 \xi_1 (
  \xi_1 \xi'_2 \xi'_2 \xi_3 \xi'_3
+ \xi'_1 \xi_4 \xi'_4 (\xi'_2 (\xi_3 \xi'_5 \xi_6+\xi'_3 \xi_5 \xi'_6) - \xi_2 \xi_5 \xi_6)
  ) \\ \nonumber
  &+&\xi_1 \xi_2 (
   \xi'_1 \xi_4 \xi'_4 (-\xi_5 \xi_6 + \xi'_5 \xi'_6)
  +(\xi_1 \xi'_2 \xi_3 + \xi'_1 \xi_4 \xi'_4 \xi_5)
  +(\xi_1 \xi'_2 \xi'_3 + \xi'_1 \xi_4 \xi'_4 \xi_6)
  ) \\ \nonumber
  &+&\xi'_1 \xi_4 (
     \xi_1 \xi_5 (\xi_2+\xi'_2 \xi'_3) (\xi_4 \xi'_5 + \xi'_4 \xi_6)
   + \xi_1 \xi'_2 \xi_3 \xi'_5 (\xi_4 \xi_5+\xi'_4 \xi'_6)
   + \xi'_1 \xi_4 \xi'_4 \xi_5 \xi'_5)\\ \nonumber
  &+&\xi'_1 \xi'_4 (
   \xi_1 \xi_6 (\xi_2+\xi'_2 \xi_3) (\xi_4 \xi_5 + \xi'_4 \xi'_6)
 + \xi_1 \xi'_2 \xi'_3 \xi'_6 (\xi_4 \xi'_5 + \xi'_4 \xi_6)
 + \xi'_1 \xi_4 \xi'_4 \xi_6 \xi'_6)
)),
\end{eqnarray}
with $\xi'_i=1 - \xi_i$.
Note that the integration domain for three-loop selfenergy diagram, $I_{3loop}$,  
can be changed by the similar way.

\par
In order to evaluate $I_{2loop}$ and $I_{3loop}$, 
%we used Double Exponential Formulas \cite{Mori_2005} 
we used DE \cite{Mori_2005} 
iteratively for the multi-dimensional integration.
Computation time for the numerical integration depends on the total number of 
evaluation points, $N_{total}$. 
In order to take advantage of the hardware characteristics of GRAPE9-MPX,
we divide $N_{total}$ into two parts, $i$-loop (outer loop) part 
and $j$-loop (inner loop) part, so as to satisfy $N_{total}= N_{i} \times N_{j}$.
This is realized by {\it loop fusion} technique, 
in which
the outer 3-dimensional integration in Eq. (\ref{eq:I2lbc}) can be performed 
in {\it i}-loop and the most inner 3-dimensions in {\it j}-loop.
In the performance measurements, we set $N_{i}$ to be $2^{18}$ 
for two-loop crossed box and $2^{23}$ three-loop selfenergy, respectively. 
On the other hand, we vary $N_j = 2^{k}$ where $k =4, 5, 6, ..., 13$.
Hereafter, we call the maximum problem size for $N_j = 2^{13}$. 
In the maximum problem size, $N_{total}$ is $2^{31}$ for the two-loop crossed box,
and $2^{36}$ for the three-loop selfenergy, respectively. 

\par
The numerical results in quadruple precision obtained by GRAPE9-MPX 
in the maximum problem size are 
$I_{2loop} = 0.008536$ and $I_{3loop} = 0.279609$, 
%$I_{2loop} = -0.0853513981$ and $I_{3loop} = 0.2796089232826$, 
%
%2014/10/8
%Below are from log files of  (2-loops) and 049perf(3-loops);
%
%
%I_{2loop, HW} = 8.5363085703599620e-02, with paramters 64 64 64 16 16 32
%I_{3loop, HW} = 2.79609276301838426987948516124359e-01, with parameters 32 32 32 32 8 8 32 32
%
%which are in good agreements with the previous works\cite{Yuasa_2012,Laporta_2000}.
which are within the relative errors of $10^{-3}$ for the two-loop crossed box and 
$10^{-5}$ for the three-loop selfenergy compared to the previous works \cite{Yuasa_2012,Laporta_2000}.
It seems that the error is slightly larger because of smaller number of $N_{total}$, 
but the calculation with this error level is achieved in a very short time, as seen later.
%
%
%
%%%%%%%%%%%%%%%%%%%%%%%%%%%%%%%%%%%%%%%%%%%%%%%%%%%%%%%%%%%%%%%%%%%%%%%%%%%%%%%%%%%%%%
\section{Performance results}

\begin{figure}[htbp]
 \begin{minipage}{0.5\hsize}
  \begin{center}
   \includegraphics[width=80mm]{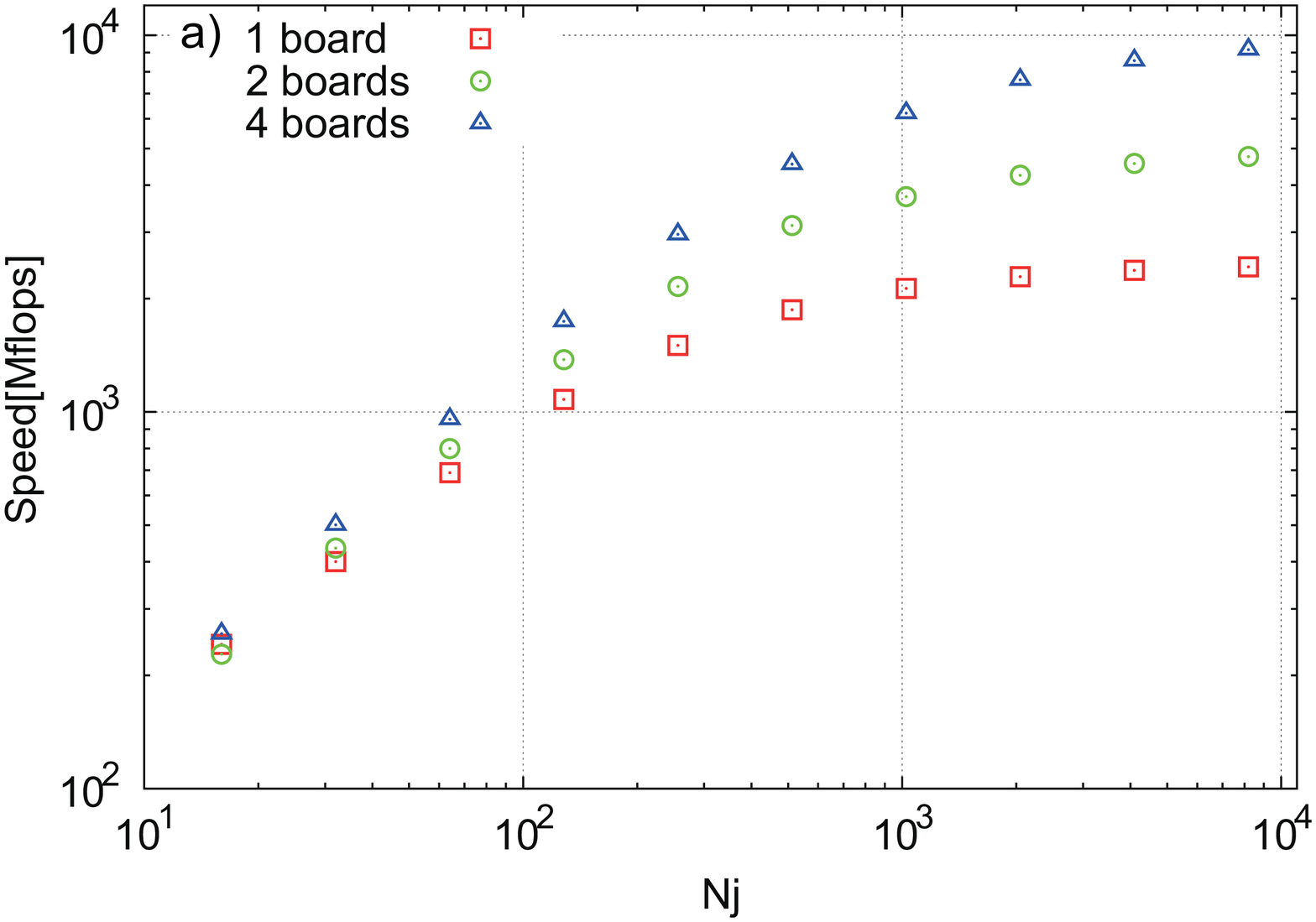}
  \end{center}
 \end{minipage}
\hfill
 \begin{minipage}{0.5\hsize}
  \begin{center}
   \includegraphics[width=80mm]{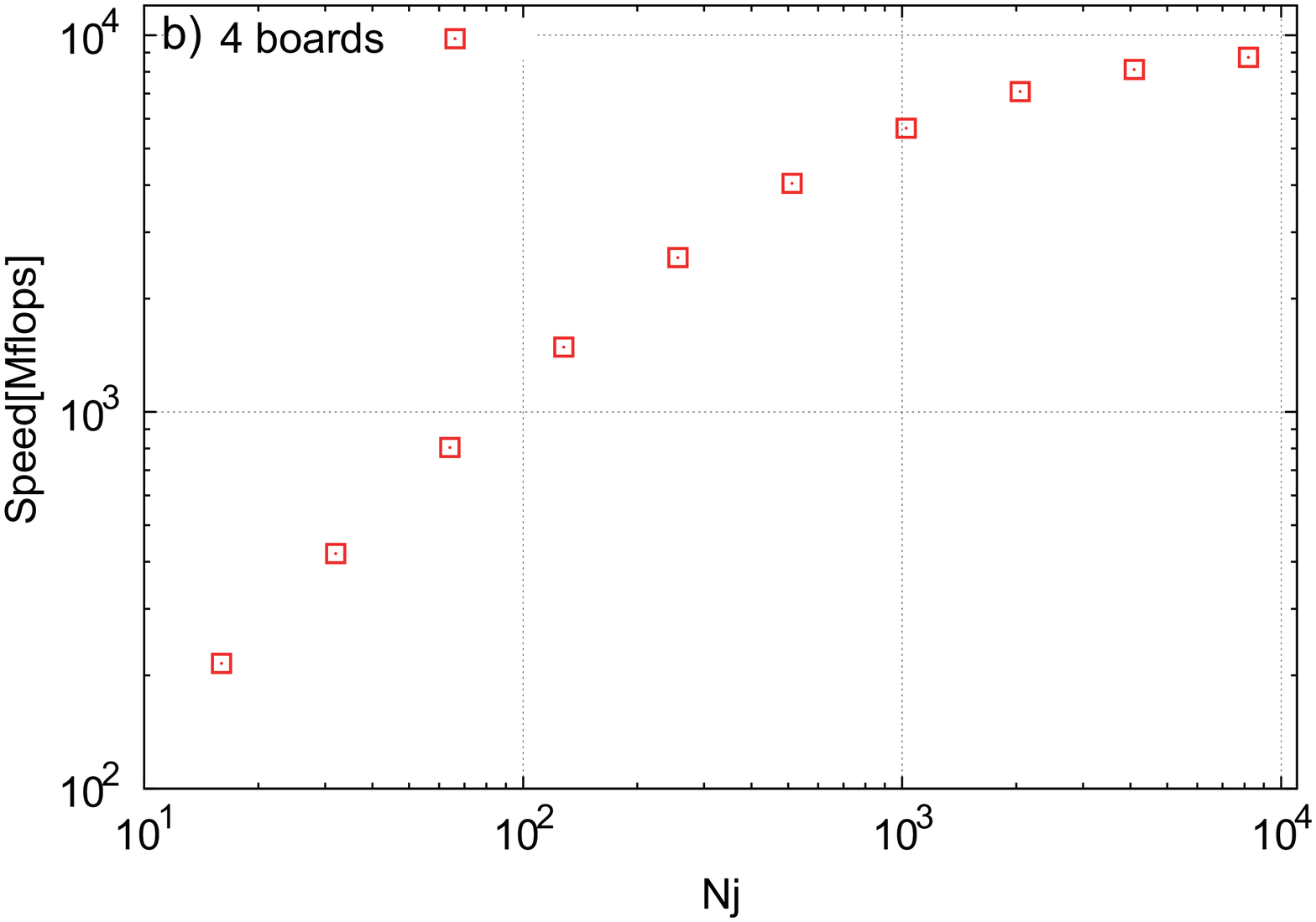}
  \end{center}
%  \caption{Performance result for three-loop selfenergy diagram}
%  \label{fig:result-3lse}
 \end{minipage}
  \caption{Performance results for two-loop crossed box diagram (left) 
   and three-loop selfenergy diagram (right).}
  \label{fig:result-2lb-3lse}
\end{figure}

We measured the performance of GRAPE9-MPX with up to 4 FPGA boards. 
It is connected a Linux PC (Scientific Linux 6.5) with Intel Xeon 2687W (3.4GHz) CPU on 
ASRock Xtreme11 motherboard (X79 chipset). 
The performance results are plotted in Fig.\ref{fig:result-2lb-3lse}.
From this figure, we can see that 
for the two-loop crossed box integral, 
the effective performance in the maximum problem size was 2.4 Gflops with a single board, 
4.7 Gflops with 2 boards, and 9.1 Gflops with 4 boards, respectively. 
On the other hand, the effective performance for the three-loop selfenergy integral 
in the maximum problem size was 8.7 Gflops with 4 boards.

\par
This figure also shows that the speed up by multiple boards is well scalable in 
the maximum problem size. 
It becomes 3.8 times faster using 4 boards,  and 1.9 times faster using 2 boards than using a single board.
On the other hand, 
the scalability is not good in smaller $N_{j}$. 
This is because the overhead by communication between the GRAPE9-MPX boards
and the host computer via PCI express is not negligible for smaller $N_j$.

We compared the efficiency of the effective performance 
for the two-loop crossed box integral and for the three-loop selfenergy integral. 
We show that 
it is about 34\% of the theoretical peak performance for the two-loop crossed box,  
whereas about 33\% for the three-loop selfenergy.  
The reason why the latter is slightly lower than the former comes 
from the difference of the number of instructions. 
From LSUMP, the operation count of the instruction is 
91 for two-loop crossed box, whereas it is 
89 for three-loop selfenergy.

%%%%%%%%%%%%%%%%%%%%%%%%%%%%%%%%%%%%%%%%%
%%%%%%%%%%%%%%%%%%%%%%%%%%%%%%%%%%%%%%%%%

We also compared the computation time for 
the two-loop integral in the maximum problem size 
measured in GRAPE9-MPX of MP4 with 4boards, 
and the time measured in software implementations 
with the same numerical format (15-bit exponent and 112-bit mantissa). 
The measured calculation time is 
21.3 sec for GRAPE9-MPX, 
1074.7 sec for {\tt quadmath} on gcc4.6.3, and 
6086.7 sec for {\tt GMP-MPFR}.  
Although we used {\tt OpenMP} for the multi-threaded 
computing and the number of threads used was 16, 
GRAPE9-MPX system was 
52.9 times faster than {\tt quadmath} and,
319.5 times faster than {\tt GMP-MPFR}. 
Thus, we could show the advantage of our system in the performance.

\section{Summary}
We have developed GRAPE9-MPX system,
an accelerator system for multi-precision arithmetic operations  
which can accelerate the calculation of Feynman loop integrals. 
Our system consists of multiple FPGA boards in which a lot of PEs are implemented 
and forms SIMD way. 
Each PE has a dedicated logic for multi-precision operations.
We measured the performance of the system in quadruple precision 
(15-bit exponent and 112-bit mantissa)  
for the two-loop crossed box and the three-loop selfenergy integrals,  
and got the effective performance of 9.1 Gflops and of 8.7 Gflops 
with 4 boards in the maximum program size, respectively.

One of the advantages of our system is that we have developed 
not only the hardware with high performance in multi-precision calculation,
but also the programming interface which enables us to port applications easily 
for our system.
The combination of Goose and LSUMP provides a simple way to use GRAPE9-MPX.
All we have to do is to insert the directive in the original code.
This simple programming model plays an essential role in treating various kinds 
of loop integrals in the automatic computation of Feynman diagrams.

Although we mainly presented the results of MP4 in this paper, 
we have measured the performance of MP6 and MP8 for the same Feynman loop integral.
For the case of the two-loop crossed box, 
we achieved 4.4 Gflops for MP6 and 2.3 Gflops for MP8 with 4 boards, 
respectively.  
The detail of the results will appear in near future.
%
%
%
%%%%%%%%%%%%%%%%%%%%%%%%%%%%%%%%%%%%%%%%%%%%%%%%%%%%%%%%%%%%%%%%%%%%%%%%%%%%%%%%%%%%%%
\section*{Acknowledgment}
We acknowledge the support from Grants-in-Aid for Scientific Research 23540328 and 24540292.
%
%
%
%%%%%%%%%%%%%%%%%%%%%%%%%%%%%%%%%%%%%%%%%%%%%%%%%%%%%%%%%%%%%%%%%%%%%%%%%%%%%%%%%%%%%%
\section*{References}
%%%%%%%%%%%%%%%%%%%%%%%%%%%%%%%%%%%%%%%%%%%
%\bibliography{biblist}
\providecommand{\newblock}{}

%%%%%%%%%%%%%%%%%%%%%%%%%%%%%%%%%%%%%%%%%%%

\end{document}